%% file: monopole.tex
\documentclass[aps,prd,preprint,groupedaddress,amssymb]{revtex4}

\usepackage{color}
\usepackage{epsfig}

\def\plb#1{Phys.~Lett.~{\bf B#1}}
\def\npb#1{Nucl.~Phys.~{\bf B#1}}
\def\prl#1{Phys.~Rev.~Lett.~{\bf #1}}
\def\prd#1{Phys.~Rev.~{\bf D#1}}

\begin{document}

\preprint{ OHSTPY-HEP-T-02-005}
\preprint{OSU-HEP-02-08}

\title{Magnetic Monopoles with Wilson loops on a 5D Orbifold}

\author{Radovan Derm\' \i\v sek}
\email[]{dermisek@pacific.mps.ohio-state.edu}

\author{Stuart Raby}
\email[]{raby@pacific.mps.ohio-state.edu}

\affiliation{Department of Physics, The Ohio State University, 174 W 18th Ave, Columbus, OH 43210}

\author{S. Nandi}
\email[]{shaown@okstate.edu}

\affiliation{Department of Physics, Oklahoma State University, Stillwater, OK
74078}

\date{\today}

\begin{abstract}
We discuss magnetic monopoles in gauge theories with Wilson loops on
orbifolds.  We present a simple example in 5 dimensions with the fifth 
dimension compactified on an $S^1/Z_2$ orbifold.   The Wilson loop in 
this $SO(3)$ example replaces the adjoint Higgs scalar (needed to break 
$SO(3)$ to $U(1)$) in the well-known 't Hooft - Polyakov construction.   
Our solution is a magnetic monopole string with finite energy, and length 
equal to the size of the extra dimension. 
\end{abstract}


\maketitle

\section{Introduction \label{sec:intro}}

Recently there has been great interest in non-abelian gauge field theories in 
4 + d dimensions with d extra dimensions compactified on an orbifold
\cite{guts,kawamura,so10,nandi,su3_unification}.
The extra dimensions can have inverse radii of order a few TeV, of order the
GUT scale or anything in between.   In most recent papers gauge and 
supersymmetry breaking, via boundary conditions imposed upon orbifold compactification, has been an interesting alternative to the traditional
Higgs mechanism.   As an illustrative and simple
example,  $S^1/ Z_2$ and $S^1/ (Z_2 \times Z_2^\prime)$ in one extra dimension
have been used to break the GUT groups $SU(5) \rightarrow SU(3) \times SU(2) \times U(1)$ \cite{guts,kawamura}, $SO(10) \rightarrow SU(4) \times SU(2)_L \times SU(2)_R$ \cite{so10}, the left-right gauge symmetry $SU(2)_L \times SU(2)_R \times
U(1)_{B-L} \rightarrow SU(2)_L \times U(1)_R \times U(1)_{B-L}$ \cite{nandi}
or the electroweak unified group $SU(3) \rightarrow SU(2) \times
U(1)$ \cite{su3_unification}.   In this paper we argue that magnetic 
monopoles are generic consequences of gauge symmetry breaking with
Wilson loops on $S^1/Z_2$ orbifolds.  The proof is by construction.  We show how to construct magnetic monopole (or more precisely, magnetic monopole string) solutions in these theories.  We also elucidate the correspondence between
compactification with Wilson loops on $S^1/Z_2$ and gauge symmetry breaking
on $S^1/ (Z_2 \times Z_2^\prime)$ orbifolds.

Recall that Dirac \cite{dirac} showed how to construct magnetic monopoles in $U(1)$ gauge theories.  The Dirac monopole is singular at the origin and thus has infinite mass.   In addition a magnetic flux tube (the so-called Dirac string) extends from the origin to spatial infinity.  Dirac showed however 
that if and only if the monopole charge $g$ satisfies the quantization condition $ g \; e = n/2 $, with $n \in \mathbb Z$ and $e$ the minimal 
electric charge, will the flux tube be an unobservable gauge artifact.

't Hooft and Polyakov \cite{'thooft_polyakov} embedded the Dirac 
monopole into a non-abelian gauge theory,
in particular Georgi--Glashow $SO(3)$ \cite{georgi_glashow}.  In this 
example $SO(3)$ is spontaneously broken to $U(1)_{EM}$ via the vacuum
expectation value [vev] ($V$) of a Higgs scalar in the
adjoint representation.  In the 't Hooft -- Polyakov construction the 
monopole singularity is removed and the Dirac string is eliminated, 
resulting in a monopole with finite mass 
$\sim 4 \pi V/e$ and two units of Dirac magnetic
charge with $g = 1/e$.   Note, the 't Hooft -- Polyakov monopole can be 
embedded into any $SU(N)$ gauge theory, see for example \cite{goldhaber_wilkinson}.

In gauge theories defined on  $M \times \Gamma$ (with $M$ a four
dimensional Minkowski space and $\Gamma$ a compact $d$ dimensional 
manifold with non-trivial homotopy $\Pi_1(\Gamma)$),
Hosotani \cite{wilson_line} showed that Wilson loops, i.e.
$\exp (i \oint A_\alpha dx^\alpha)$ integrated around a non-contractible 
closed loop in $\Gamma$, can spontaneously break the gauge symmetry.   
The Wilson loop acts like the vev of a Higgs scalar in the adjoint representation.   Wen and
Witten \cite{wen_witten} considered magnetic monopoles with Wilson loops
on compact manifolds.  They assumed $\Pi_1(\Gamma) = Z_n$ and  
discussed the allowed values of magnetic monopole charges in such theories.
A simple explicit example of a monopole construction with Wilson loops was 
discussed by Lee et al. \cite{lee_etal} using an $SU(3)$ gauge theory
defined on $M \times S^3/Z_2$.  In their
example, the Wilson loop breaks $SU(3) \rightarrow SU(2) \times U(1)$.

In this paper, we extend the construction of magnetic monopoles to 
orbifolds \footnote{These are actually monopole strings since they have
constant energy density and charge in the fifth direction.}.
In particular we consider the simple example of an $SO(3)$ gauge theory 
defined on the orbifold $M \times S^1/Z_2$ with a background gauge
field, or equivalently, 
$M \times S^1/(Z_2 \times Z_2^\prime)$.  Our discussion is self-contained,
however for a recent discussion of Wilson loops on orbifolds see \cite{hall_etal,kubo}.  Finally our analysis is easily extended to any $SU(N)$ gauge group defined on $M \times S^1/Z_2$ using previous analyses for the extension of 't Hooft -- Polyakov monopoles  \cite{goldhaber_wilkinson}.

The paper is organized as follows.   In section \ref{sec:so3in5d}
we introduce the Wilson loop symmetry breaking mechanism in the simple
example of the circle $S^1$.  We then generalize this discussion
in section \ref{sec:orbifold} to the orbifold $S^1/Z_2$.  In \ref{sec:equivalence} we elucidate the equivalence of gauge symmetry breaking with Wilson loops on $S^1/Z_2$ and gauge symmetry breaking on the orbifold $S^1/(Z_2 \times Z_2^\prime)$.   In section \ref{sec:monopoles}
we explicitly construct the monopole string solution and discuss some
of its properties.   Finally in \ref{sec:conclusion} we summarize
our results and consider possible phenomenological ramifications. 

\section{SO(3) gauge theory on $M \times S^1$ \label{sec:so3in5d}}

Consider a general gauge theory with symmetry group $G$ in five
dimensional spacetime.   The Lagrangian is given by
\begin{equation}
{\cal L}_5 = - \frac{1}{4e_5^2 k} Tr(F_{M N} F^{M N})
\label{eq:lag1}
\end{equation}
where $F_{M N} \; \equiv \; \sum_a F^a_{M N} T^a$,  $\;\; T^a$ are generators in some finite dimensional representation of $G$ normalized such that
$Tr (T^a T^b) = k \delta^{a b}$ and $M, N = \{ 0, 1, 2, 3, 5 \}$:
\begin{equation}
F_{M N} = \partial_M A_N - \partial_N A_M + i [ A_M, A_N ].
\end{equation}
(For the adjoint representation of $SO(3)$ we use the standard normalization
of the generators with $k = 2$.)
The gauge transformation of the gauge field $A_{M} (x_\mu, y) \; \equiv \; \sum_a A_M^a T^a (x_\mu, y)$ (greek indices correspond to 4-dimensional Minkowski 
spacetime and $y \ \equiv \ x_5$) is given by
\begin{equation}
A_M (x_\mu, y) \rightarrow U A_M (x_\mu, y) U^{\dagger} - i U \partial_M U^{\dagger},
\label{eq:gauge_tr}
\end{equation}
where 
\begin{equation}
U = \exp (i \theta^a (x_\mu, y) T^a).
\label{eq:gauge_tr2}
\end{equation}

In our notation, Eq. (\ref{eq:lag1}), the gauge fields have mass dimensions [1],  and the charge $e_5$ has dimension [-1/2].  We can also define the effective four dimensional, dimensionless, gauge coupling $e$ by rescaling $e_5$ in Eq. (\ref{eq:lag1}) via the expression $e_5 = \sqrt{2 \pi R} \; e$.  Note, if $\partial_5 A_\mu = 0$, then 
$F_{\mu 5}$ reduces to the covariant derivative of the 5th component of the gauge field $A_5$. In this case we can conveniently define $\Phi \equiv A_5/e_5
= \tilde \Phi /\sqrt{2 \pi R}$, where the scalars $\Phi$ and  $\tilde \Phi$ have dimension $[3/2]$ and $[1]$.  The Lagrangian (\ref{eq:lag1}) can then be rewritten in the suggestive form:
\begin{equation}
{\cal L}_5 = \frac{1}{2 \pi R} \left[ - \frac{1}{4e^2 k} Tr( F_{\mu \nu} F^{\mu \nu}) + \frac{1}{2 k} Tr (D_\mu \tilde \Phi D^\mu \tilde \Phi) \right] .
\label{eq:lag2}
\end{equation}
This resembles the Georgi--Glashow model \cite{georgi_glashow} of an $SO(3)$ gauge theory interacting with an isovector Higgs field. There are two differences, however. First, there is no  
potential $V(\tilde \Phi) = \lambda (\tilde \Phi^a \tilde \Phi^a - V^2)^2$ for the Higgs field which would break the gauge symmetry down to 
$U(1)$ and second, the Higgs field depends on the 5th coordinate.
Although this analysis is limited to gauge fields 
satisfying $\partial_5 A_\mu = 0$, it nevertheless inspires the following discussion of symmetry breaking via Wilson loops and the further 
consideration of monopoles with Wilson loops.  In general, however,
$\partial_5 A_\mu
\neq 0$ and we need to keep the full $Tr (F_{\mu 5}^2)$ term.

\subsection{Wilson loop gauge symmetry breaking on $ M \times S^1$ \label{sec:wilson-line1}}

Assume the 5th dimension is compactified on a circle $S^1$
parametrized by $y \in [0, 2\pi R]$.  The gauge symmetry can then be broken 
by the presence of a background gauge field $A_5$.  This symmetry breaking
mechanism is known as Hosotani or Wilson loop symmetry breaking \cite{wilson_line}. 
Consider the constant background to be along the third isospin direction,
\begin{equation}
A_5 (y) = A^3_{5} T^3. 
\label{eq:a5}
\end{equation}
Using the single valued gauge transformation (periodic under $y \rightarrow y + 2 \pi R$) given by 
Eqs.~(\ref{eq:gauge_tr},\ref{eq:gauge_tr2}) with $\theta (x_\mu, y) = - n y/ R$, $n \in \mathbb Z$: 
\begin{equation}
U(y) = \exp \left(- i n T^3 \frac{y}{R} \right),
\end{equation}
we obtain the transformation of $A^3_{5}$: 
\begin{equation}
A^3_{5} \rightarrow A^3_{5} + n/R. 
\label{eq:a5_gauge_tr}
\end{equation}
Therefore the gauge non-equivalent values of 
$A^3_{5}$ can be chosen to lie between $0$ and $1/R$. The holonomy due to this constant background gauge field is given by
\begin{equation}
T = \exp \left( i \oint A_5 dy \right) = \exp \left( i \alpha T^3 \right).
\end{equation}
with the arbitrary parameter $\alpha \equiv 2 \pi R A_5^3$.  
Note the set of possible holonomies $\{ \openone, T^{\pm 1}, T^{\pm 2}, \cdots \}$ provides a mapping of the gauge group into the discrete group $\mathbb Z$.
This non-trivial holonomy affects the spectrum of the theory. 
A massless periodic scalar field $\phi$ (satisfying $\phi(y + 2\pi R) = \phi(y)$) with isospin eigenvalue $I_3$ can be decomposed into 
Kaluza-Klein modes 
\begin{equation}
\phi_{(n)}(x_\mu) \exp (iny/R) .
\end{equation}
The 5-dimensional wave equation $D^M D_M \phi = 0$ splits 
into an infinite set of 4-dimensional wave equations for Kaluza-Klein modes $\phi_{(n)}$ with masses given by 
\begin{equation}
m^2_{(n)} \phi_{(n)} \exp (iny/R) = - \left( \partial_y + i A^3_{5} T^3 \right)^2 \phi_{(n)} \exp (iny/R)
= \left( \frac{n}{R} + A_5^3 I_3 \right)^2 \phi_{(n)} \exp (iny/R). 
\label{eq:KK}
\end{equation}
It is now easy to obtain the spectrum of gauge fields \footnote{Consider a background field gauge with $A_M = B_M + a_M$ where $B_M$ is the background value of the gauge field and $a_M$ are the small fluctuations. The background
covariant derivative is given by $D_M \equiv \partial_M + i [ B_M , \; ]$.
If we use the covariant gauge fixing condition $D^M a_M \equiv 0$, then the gauge field equations of motion are given by 
$D^M D_M a_N + 3 i [B_{N M}, a^M] = 0$. Note, for a constant background
gauge field $B_{N M} \equiv 0$.}.  The gauge field $A^3_{\mu}(y)$ has $I_3 = 0$ and therefore its 
KK modes 
are not affected by the holonomy. The zero mode of this field corresponds to the gauge field of the unbroken $U(1)$.   On the other hand,
the masses of the KK modes of the $W^\pm$ gauge bosons, with $I_3 = \pm 1$,
are given by $m_{(n)} = | \frac{n}{R} \pm A_5^3 |$. If $A_5^3 \neq 
\frac{k}{R}$, where $k \in \mathbb Z$, the gauge bosons $W^\pm$ are all massive. Clearly the $SO(3)$ symmetry is broken to $U(1)$.  Note, the
symmetry breaking scale satisfies $0 \leq A_5^3 < 1/R$, but is 
otherwise unconstrained. 

\subsection{Gauge Picture with Vanishing Background}

A constant background gauge field $A^3_{5}$ may be gauged away with the non-periodic gauge transformation 
\begin{equation}
U(y) = \exp \left( i y A_5^3 T^3  \right).
\label{eq:gauge_away}
\end{equation}
In this gauge the covariant derivative in Eq.~(\ref{eq:KK}) is trivial, i.e. $D_5 = \partial_5$.  Nevertheless it is easy to see that, as expected, the physics is unchanged.   

This gauge transformation is not single valued and thus the periodicity condition \newline $\phi(y + 2\pi R) = \phi(y)$ becomes
\begin{equation}
\phi(y + 2\pi R) = \exp \left( i \alpha T^3 \right) \phi(y).
\label{eq:translation}
\end{equation}
Now the mode expansions are of the form
\begin{equation}
\phi_{(n)}(x_\mu) \exp \left[i \left(n/R + A_5^3 I_3\right)y \right] 
\end{equation}
resulting in the identical spectrum as before.

\section{SO(3) gauge theory on $S^1/Z_2$}

\subsection{The $S^1/Z_2$ orbifold \label{sec:orbifold}}

The $S^1/Z_2$ orbifold is a circle $S^1$ modded out by a $Z_2$ parity symmetry: $y \rightarrow  - y$.
The 5th dimension is now a line segment $y \in [0, \pi R]$. This orbifold has two fixed points at $y = 0$ and $\pi R$.
The Lagrangian (\ref{eq:lag1})  is invariant under the parity transformation  
\begin{equation}
A_\mu ( -y ) = A_\mu ( y )
\label{eq:parity}
\end{equation}
\begin{equation}
A_5 (-y) = - A_5 (y).
\label{eq:parity5}
\end{equation}
As in the case of compactification on a circle we consider a constant 
background for $A_5^3$ (Eq. (\ref{eq:a5})).  Clearly such a background is
not consistent with the parity operation, Eq. (\ref{eq:parity5}).
However, following \cite{hall_etal} we define a generalized parity by
combining the parity transformation (\ref{eq:parity5}) with the gauge transformation (\ref{eq:a5_gauge_tr}), for $n = 1$, $A^3_5 \rightarrow A^3_5 + 1/R$.   We then look for a consistent
solution with constant $A_5^3$.  There are now only two possible values for $A^3_5$.
The possibility $A^3_5 = 0$ is obviously allowed, but in this case the gauge symmetry is unbroken. The only 
nontrivial choice corresponds to 
$A^3_5 (y)  = \frac{1}{2R}$ which changes sign under the ``naive" parity, $A^3_5 (-y) = - \frac{1}{2R}$, but is gauge equivalent to its original 
value. Therefore, instead of (\ref{eq:parity}) -- (\ref{eq:parity5}) we define 
the fields for negative $y$, in the region  $- \pi R < y < 0$, in terms of the fields defined for positive $y$ in the fundamental domain, $0 < y < \pi R$, via the generalized parity transformation (i.e. a combined  ``naive" parity transformation (\ref{eq:parity5}) and a gauge transformation) such that, in general:
\begin{equation}
A_\mu(- y) = U(-y) A_\mu (y) U^{\dagger}(-y) - i U(-y) \partial_\mu U^{\dagger}(-y), 
\label{eq:parity1}
\end{equation}
\begin{equation}
A_5(- y) = - U(-y)  A_5(y) U^{\dagger}(-y)  - i U(-y)  \partial_{-y} U^{\dagger}(-y),
\label{eq:parity2}
\end{equation}
with
\begin{equation}
U (y) = \exp \left( -  i \frac{y}{R} T^3 \right) 
\label{eq:gauge_tr_t3}
\end{equation}

It is useful to define new fields, $W^\pm$, in a usual way from $A^1$ and $A^2$: 
\begin{equation}
W^\pm = \frac{1}{\sqrt{2}} \left( A^1 \mp i A^2  \right) , \quad 
T^\pm = \frac{1}{\sqrt{2}} \left( T^1 \pm i T^2  \right).
\end{equation}
With this definition we have $A^1 T^1 + A^2 T^2 = W^+ T^+ + W^- T^-$ and $[T^3, T^\pm ] = \pm T^\pm$.
Using  the identity
\begin{equation}
 \exp \left( i \frac{y}{R} T^3 \right) T^\pm \exp \left(- i \frac{y}{R} T^3 \right) = 
\exp \left(\pm i \frac{y}{R} \right) T^\pm
\end{equation} 
it is easy to show that the generalized 
parity tranformation acts on gauge fields as follows:
\begin{eqnarray}
W^\pm_\mu (-y) &=& \exp \left( \pm i \frac{y}{R} \right) W^\pm_\mu (y) , \label{eq:gen_par1}\\
W^\pm_5 (-y) &=& - \exp \left( \pm i \frac{y}{R} \right) W^\pm_5 (y) ,\label{eq:gen_par2}\\
A^3_\mu (-y) &=& A^3_\mu (y),\label{eq:gen_par3} \\
A^3_5 (-y) &=& - A^3_5 (y) + \frac{1}{R}. \label{eq:gen_par4}
\end{eqnarray}

To summarize, using a more compact notation, we have the following constraints on the fields (valid for all modes, except the constant piece of $A_5^3$).  Under the generalized parity transformation 
the fields $\phi_P$ (with $P = \pm 1$) satisfy:
\begin{equation}
\phi_P (-y) = P \exp \left(i \frac{y}{R} I_3 \right) \phi_P (y) 
\end{equation}
with isospin
 eigenvalue $I_3 = \pm 1, 0$.   The periodicity condition is given by: 
\begin{equation}
\phi_P (y + 2 \pi R) = \phi_P (y). 
\end{equation}
We then obtain the following decomposition into KK modes:
\begin{eqnarray}
\phi_+ (x_\mu , y) = \sum^\infty_{n = 0} \phi_+^{(n)} (x_\mu) \, \exp \left( -i \frac{y}{2R} I_3 \right) \, \cos n
\frac{y}{R} \quad &{\rm for}& {\rm even} \;\; I_3, \label{eq:phi1} \\
\phi_+ (x_\mu , y) = \sum^\infty_{n = 0} \phi_+^{(n)} (x_\mu) \, \exp \left( -i \frac{y}{2R} I_3 \right) \, \cos (n+ 1/2)
\frac{y}{R}  \quad &{\rm for}& {\rm odd} \;\; I_3, \label{eq:phi2} \\
\phi_- (x_\mu , y) = \sum^\infty_{n = 0} \phi_+^{(n)} (x_\mu) \, \exp \left( -i \frac{y}{2R} I_3 \right) \, \sin (n+ 1)
\frac{y}{R}   \quad &{\rm for}& {\rm even} \;\; I_3, \label{eq:phi3}\\
\phi_- (x_\mu , y) = \sum^\infty_{n = 0} \phi_+^{(n)} (x_\mu) \, \exp \left( -i \frac{y}{2R} I_3 \right) \, \sin (n+ 1/2)
\frac{y}{R}  \quad &{\rm for}& {\rm odd} \;\;  I_3. \label{eq:phi4}
\end{eqnarray}
From transformations (\ref{eq:gen_par1}) -- (\ref{eq:gen_par4}) we see that the KK mode expansion of 
$A^3_\mu$ [(+) field with $I_3=0$] is given in Eq.~(\ref{eq:phi1}) with corresponding masses $n/R$. This is the 
only field which has a 
zero mode. It corresponds to the 
gauge field of the unbroken $U(1)$. The expansion of 
$W^\pm_\mu$ [(+) field with $I_3=\pm 1$] is given in  
Eq.~(\ref{eq:phi2}) with corresponding masses $(n + 1/2)/R$. Similarly, the expansion of $W^\pm_5$ [(--) field 
with $I_3=\pm 1$] is given in
 Eq.~(\ref{eq:phi4}) with corresponding masses $(n + 1/2)/R$. And finally, the expansion of $A^3_5$ 
[(--) field with $I_3=0$] is given by 
Eq.~(\ref{eq:phi3}) up to the value of the constant background:
\begin{equation}
A^3_5 (x_\mu , y) = \frac{1}{2R} \;+\; \sum^\infty_{n = 0} A_5^{3 (n)} (x_\mu)  \, \sin (n+ 1) \frac{y}{R}. 
\end{equation}

The holonomy $T$ in this case is given by 
\begin{equation}
T = \exp ( i \oint A_5^3 T^3 ) = \exp (i \pi T^3 ) = diag ( -1, -1, 1). 
\label{eq:holonomy1}
\end{equation}
 Hence $T^2 = \openone$ or the set of possible holonomies $\{ \openone, T \}$ maps the gauge group into the discrete group $\mathbb Z_2$.
Unlike the case of Wilson loops on $S^1$ discussed in section 
\ref{sec:so3in5d}, the background gauge field and consequently
the holonomy on $S^1/Z_2$ can only take discrete values.

Now let us consider the gauge picture with vanishing background gauge field. As in the case of compactification on a circle, we can gauge away the constant background by the non-single valued gauge transformation given in Eq.~(\ref{eq:gauge_away}). 
The transformations under the generalized parity are now those of Eqs.~(\ref{eq:parity}) and (\ref{eq:parity5}). In addition the  
non-single valued gauge transformation changes the periodicity condition as in Eq.~(\ref{eq:translation}) with $\alpha = \pi$.
 
To obtain the spectrum of KK modes of a field $\phi$ we consider both the transformation under parity and the effect of a non-trivial holonomy. 
Under parity, 
\begin{equation}
{\cal P}: \quad \phi_{P T}(y) \; \rightarrow \; \phi_{P T} (-y) = P \phi_{P T} (y),
\end{equation}
with $P^2 = 1$ or $P = \pm 1$. When going around the circle, the fields transform in the following way:
\begin{equation}
{\cal T}: \quad \phi_{P T} (y) \; \rightarrow \; \phi_{P T} (y + 2 \pi R) = T \phi_{P T} (y)
\label{eq:holonomy}  
\end{equation} 
with $T^2 = \openone$ or $T = \pm 1$. 
Therefore there are four different kinds of fields $\phi_{\pm \pm}$ corresponding to the four different combinations of  
$(P,T)$.  
It is easy to see that a field with given $(P,T)$ can be expanded 
into the following modes: 
\begin{eqnarray} 
&\xi_n (+,+) &= \ \cos n \frac{y}{R} \nonumber \\
&\xi_n (+,-)&= \ \cos (n + 1/2)\frac{y}{R} \nonumber \\
&\xi_n (-,+)&= \ \sin (n + 1) \frac{y}{R} \nonumber \\
&\xi_n (-,-)&= \ \sin (n + 1/2) \frac{y}{R} 
\end{eqnarray}
Only the $(+,+)$ fields have massless zero modes. 
Of all the gauge fields only $A^3_\mu$ is a $(+,+)$ field with a zero mode. $W^\pm_\mu$, $A^3_5$ and $W^\pm_5$  are $(+,-)$, $(-,+)$ and $(-,-)$ fields
, respectively.  Clearly the mode expansion and the corresponding KK 
masses are the same as in the previous picture.  Note, our gauge transformation
parameters (Eq. (\ref{eq:gauge_tr2})) are constrained to satisfy $\theta^3(x_\mu, y) = \theta^3_n (x_\mu) \xi_n (+,+)$  and
$\theta^{1, 2}(x_\mu, y) = \theta^{1, 2}_n (x_\mu) \xi_n (+,-)$.  
Hence,  $SO(3)$ is the symmetry
everywhere in the five dimensions, EXCEPT on the boundary at $y = \pi R$.

\subsection{Correspondence to $S^1/(Z_2 \times Z_2^\prime )$ orbifold
\label{sec:equivalence}}

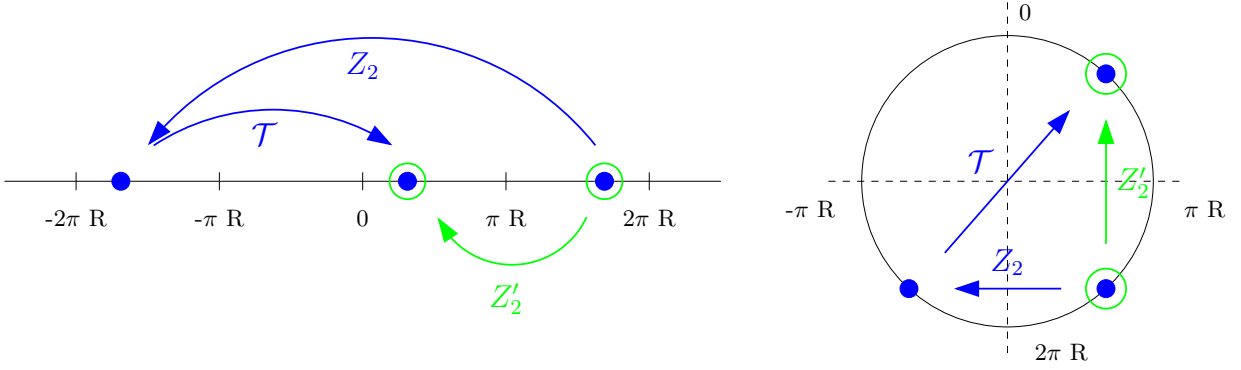
\begin{figure}[t]
\input{line_circle.pstex_t}
\caption{\label{fig:line}The $Z_2^\prime$ parity transformation is equivalent to the combined $Z_2$ parity transformation 
and translation ${\cal T}$.} 
\end{figure}

The $S^1/Z_2$ orbifold with holonomy $T$ in the gauge picture without a
constant background gauge field is directly related to the $S^1/(Z_2 \times Z_2^\prime )$ orbifold used recently in the literature \cite{guts,kawamura,so10,nandi,su3_unification}.   This correspondence is
also evident in the work of Ref. \cite{hall_etal,scherk_schwarz}.
We just need to identify the $S^1/(Z_2 \times Z_2^\prime )$ orbifold 
with $S^1$, a circle of circumference $4 \pi R$, divided by the $Z_2$ transformation $y \rightarrow -y$ and $Z_2^\prime$ transformation $y^\prime \rightarrow - y^\prime$, where $y^\prime \equiv y - \pi R$. The physical space is again the line segment $y \in [0, \pi R]$ with orbifold 
fixed points at $y = 0$ and $ \pi R$. It is easy to see that ${\cal P}^\prime \in Z_2^\prime$ in this picture corresponds to the combined 
translation and parity transformation in the previous picture, namely 
${\cal P}^\prime = {\cal T} {\cal P}$.  Note, a point at $y = y_0$ which corresponds to $y^\prime = y_0 - \pi R$ is transformed by $Z_2^\prime$ into the 
point $y^\prime = - (y_0 - \pi R)$ corresponding to $y = - y_0 + 2 \pi R$; this is equivalent to the action of 
${\cal T} Z_2$ on the point at $y = y_0$, see Fig.~\ref{fig:line}.

The action of $Z_2$ on the fields is given by
\begin{equation}
{\cal P}: \quad \phi_{P P^\prime}(y) \; \rightarrow \; \phi_{P P^\prime} (-y) = P \phi_{P P^\prime} (y),
\end{equation}
with $P^2 = 1$ or $P = \pm 1$. 
Similarly, under $Z_2^\prime$ we have
\begin{equation}
{\cal P^\prime}: \quad \phi_{P P^\prime} (\pi R + y^\prime) \; \rightarrow \; \phi_{P P^\prime} (\pi R - y^\prime) = P^\prime    \phi_{P P^\prime} (\pi R + y^\prime) 
\label{eq:holonomy2}  
\end{equation} 
with $P^\prime = T P$ and $(P^\prime)^2 = \openone$ or $P^\prime = \pm 1$.

It is easy to see what the holonomy means in this picture. Since points $y_0$ and $y_0 + 2 \pi R$ are identified, 
the closed loop corresponds to going around half of the circle (the circumference of the circle in this picture is
$4 \pi R$). Going around the whole circle (from $y_0$ to $y_0 + 2 \pi R$ and then from $y_0 + 2 \pi R$ to $y_0 + 4 \pi 
R$) clearly corresponds to $T^2$.  From Eq. (\ref{eq:parity5}) we see that going from $y_0 + 2 \pi R$ to $y_0 + 4 \pi
R$ is equivalent to going backwards from $y_0 + 2 \pi R$ to $y_0$. Therefore $T^2 = \openone$ and there are only two 
possibilities for holonomy, $T = +1$ and $T= -1$, the same as in the $S^1/Z_2$ picture.   Hence we have $T \in \mathbb Z_2$.  Note, in the above we have assumed that $P$ and $T$ can be simultaneously diagonalized.  In general however
$P$ and $T$ do not commute.  In this case we would have $P \ T \ P = T^{-1}$.

\subsection{Monopole string on $S^1/Z_2$ \label{sec:monopoles}}

We saw that the gauge theory in 5-dimensions becomes a ``gauge - Higgs" theory after the 5th dimension is compactified.
The Higgs potential which breaks the $SO(3)$ gauge symmetry to $U(1)$ is absent, however its effect can be replaced 
by the Wilson loop along the compactified dimension. It was shown by 't Hooft and Polyakov \cite{'thooft_polyakov} that 
the Georgi--Glashow model has a magnetic monopole solution to the equations of motion. 
It is natural to ask whether magnetic monopoles are present in the compactified 5-dimensional gauge theory and what is the 
correspondence with the usual 't Hooft -- Polyakov solution.

The equations of motion corresponding to the Lagrangian~(\ref{eq:lag2}) are:
\begin{equation}
D_\mu D^\mu \tilde \Phi = 0 \;\;, \quad D_\nu F^{\mu \nu} = i e^2 [\tilde \Phi, D^\mu \tilde \Phi] .
\label{eq:eq_of_motion}
\end{equation}
They correspond to the equations of motion of the Georgi--Glashow model in the absence of the Higgs potential.

Consider the ansatz (for $ 0 < y < \pi R$):
\begin{equation}
A_5/e \equiv \tilde \Phi = \frac{1}{2 R e} (\hat{\roarrow{r}} \cdot \roarrow{T}) \ F(r) \; , \\
\label{eq:ansatz1}
\end{equation}
\begin{equation}
A_{i} = - \frac{1}{r} (\roarrow{T} \times \hat{\roarrow{r}})_i \ G(r) \;,  \quad A_{0} = 0 \; , 
\label{eq:ansatz2}
\end{equation}
where $r = \sqrt {x^2_i}$, $\hat r_i = x_i/r$ and $F(r)$ and $G(r)$ are dimensionless functions.   Asymptotically, for $r \rightarrow \infty$
we have $ G(r) \rightarrow 1$.  Note, the constant
$\frac{1}{2 R}$ in the normalization of $A_5$ has been fixed by the vacuum
boundary conditions with the choice $F(r) \rightarrow 1$ as $r \rightarrow \infty$ (see discussion below).
This is exactly the 't Hooft--Polyakov
ansatz, and therefore it is a solution to the equations of motion, Eq. (\ref{eq:eq_of_motion}) with 
\begin{equation} V \equiv \lim_{r \rightarrow \infty} \sqrt{Tr (\tilde \Phi^2)/k} =  \frac{1}{2 R e} . 
\end{equation}  

In order to complete the solution we need to extend the above solution to negative $y$ (i.e.  $- \pi R <  y < 0$).    As in the case with a constant background field $A_5$ we use
the generalized parity operation, Eqs. (\ref{eq:parity1}) and (\ref{eq:parity2}),
 now with  
\begin{equation}
U = \exp \left( -  i \frac{y}{R} \hat{\roarrow{r}} \cdot \roarrow{T} \right),
\label{eq:gauge_tr_rdott}
\end{equation}
we obtain
\begin{equation}
A_5(-y)/e \equiv \tilde \Phi(-y) = \frac{-F(r) + 2}{2 R e} \ (\hat{\roarrow{r}} \cdot \roarrow{T}) \; , \\
\label{eq:ansatz3}
\end{equation}
\begin{equation}
A_{i}(-y) = - \frac{G(r) - 1 }{r} \; (\roarrow{T} \times \hat{\roarrow{r}})_i \ \cos\frac{y}{R} + \frac{G(r) - 1}{r} \; (T_i - \hat r_i (\hat{\roarrow{r}} \cdot \roarrow{T})) \ \sin\frac{y}{R} - \frac{1}{r} \ (\roarrow{T} \times \hat{\roarrow{r}})_i .
\label{eq:ansatz4}
\end{equation}

Note,  that the asymptotic values of $A_i$ and $A_5$, normalized as in Eq. (\ref{eq:ansatz1}), for $r \rightarrow \infty$ satisfy  $A_i(-y) = A_i(y)$ 
and $A_5(-y) = A_5(y)$.
Hence we obtain the asymptotic holonomy  
\begin{equation}
\lim_{r \rightarrow \infty} T(r) = \exp (i \pi \hat{\roarrow{r}} \cdot \roarrow{T}) 
\end{equation}
satisfying the condition $T^2 = \openone$, i.e. $T \in \mathbb Z_2$.   Moreover 
in any given spatial direction $\hat{\roarrow{r}}$, the asymptotic holonomy 
is gauge equivalent to the vacuum value, Eq. (\ref{eq:holonomy1}).   It 
is this physical requirement, that asymptotically far away from the 
monopole we recover the vacuum holonomy, which fixes the asymptotic 
magnitude of $A_5$, Eq. (\ref{eq:ansatz1}).    Note, in the case of a 
simple circle, discussed in section \ref{sec:wilson-line1}, 
$T \in \mathbb Z$ and the magnitude of $A_5$ is arbitrary.
In this case, the monopole mass can be taken continuously to zero.
Hence monopoles on $S^1$ are unstable.  

Although the form of the gauge fields for $- \pi R <  y < 0$,
defined by the generalized parity transformation of the 't Hooft ansatz 
for $0 < y < \pi R$ is quite complicated, it is easy to see that they 
are a solution to the field equations.  This
is because the action is both parity and gauge invariant.   In fact the
action 
\begin{equation}
S \equiv \int d^4x \int_{- \pi R}^{+ \pi R} dy  {\cal L} =
    2 \int d^4x \int_{0}^{+ \pi R} dy  {\cal L}
\end{equation}
is completely defined in terms of the fields in the fundamental domain
$0 \leq y \leq \pi R$.

The asymptotic ($r \rightarrow \infty $) gauge field strength
is given by 
\begin{equation}
F_{i j} = - \frac{\epsilon_{ijk} \; \hat r_k \ (\hat{\roarrow{r}} \cdot \roarrow{T})}{ r^2} .
\end{equation}
The asymptotic $U(1)$ abelian magnetic field is then given by
\begin{equation}
B_i \equiv  - \frac{1}{2 e k} \ \epsilon_{ijk} \ Tr \left( ( \hat{\tilde \Phi} ) \  F_{j k} \right)  =  \frac{\hat r_i}{e \; r^2}
\end{equation}
where $ \hat{\tilde \Phi} \equiv  \tilde \Phi/V$.
Therefore the solution is a magnetic monopole string with total magnetic charge $g =  1/e$ or equivalently a magnetic charge per unit length in the 5th direction given by  $g/\pi R$.

The monopole string energy density is given by
\begin{equation}
{\cal H} = \frac{1}{2 \pi R} \; \left[ \frac{1}{4 e^2 k} Tr (F_{i j} F^{i j}) + \frac{1}{2 k} Tr (D_i \tilde \Phi D_i \tilde \Phi) \right] .
\label{eq:h}
\end{equation}
It is a constant function of $y$ and thus we should really talk about a monopole
string stretched in the 5th direction from $y = 0$ to $y = \pi R$.  The energy density, Eq. (\ref{eq:h}), is the usual four dimensional energy density divided by the length of the fifth dimension and the energy per unit length of the monopole string is obtained by integrating ${\cal H}$ over the three flat spatial dimensions.  
Note, the integrated energy density from Eq. (\ref{eq:h}) can be written as
\begin{equation}
 H = \int d^3x \frac{1}{k} Tr \left[ \frac{1}{4} \left( \frac{1}{e} F_{i j} \mp \epsilon_{ijk} D_k \tilde \Phi \right)^2 \pm 
\frac{1}{2 e} \ \epsilon_{ijk} \ F_{i j} \ D_k \tilde \Phi \right]. 
\label{eq:h2}
\end{equation}
where the integration over $y$ has been performed.
The second term can be rewritten using Bianchi identity as $\frac{1}{2 e k} \epsilon_{ijk} \partial_k Tr (F_{i j} \tilde \Phi)$ 
and its contribution to the energy of the monopole is 
\begin{equation} 
\pm \frac{1}{2 e k} \ \epsilon_{ijk} \int d^3 x \, \partial_k Tr (F_{i j} \tilde \Phi) = \pm V \int \roarrow{B} \cdot d \roarrow{S} = \pm 4 \pi V g .  
\end{equation}
When the first term in (\ref{eq:h2}) vanishes the monopole solution is
said to satisfy the Bogomol'nyi bound and such monopoles are called
BPS monopoles.    In fact, the general 't Hooft -- Polyakov monopole solution
reduces to a BPS monopole in the limit that the Higgs potential for the
adjoint scalar vanishes.   Hence our monopole strings are in fact BPS 
monopole strings and their mass is given by
\begin{equation}
M_m = \frac{4 \pi V}{e} = \frac{M_W}{\alpha} = \frac{1}{2 \alpha R} 
\end{equation}
where $\alpha = e^2/ 4 \pi$ is the dimensionless fine structure constant at the scale $1/R$, and $R$ is the orbifold radius.

 It is also important to express the equations for
the BPS condition and the monopole energy density in an explicitly gauge invariant and 5D covariant form.
The BPS condition is
\begin{equation}
F_{i j} = \pm \epsilon_{i j k} F_{k 5}
\end{equation}
and the energy density is given by
 \begin{equation}
 {\cal H} =  \pm \frac{1}{2 e_5 k} \ \epsilon_{ijk} \ Tr( F_{i j} \ D_k \Phi )  =  
\pm \frac{1}{2 e_5^2 k} \epsilon_{ijk} \  Tr(F_{i j} \ F_{k 5} )  =
\pm \frac{1}{8 e_5^2 k} \epsilon_{0 N P Q R} \  Tr(F^{N P} \ F^{Q R} ).  
\label{eq:h3}
\end{equation}
Note it is then clear that the five dimensional Hamiltonian density
is the time component of a five vector given by
\begin{equation}
 {\cal P}^M = \pm \frac{1}{8 e_5^2 k} \ \epsilon^{M N P Q R} \ Tr( F_{N P} \ F_{Q R} ) \equiv \partial_N K^{M N} ,
\label{eq:P}
\end{equation}
with
\begin{equation}
 K^{M N} = \pm \frac{1}{4 e_5^2 k} \ \epsilon^{M N P Q R} \ 
Tr( A_{P} \ F_{Q R} - i \frac{2}{3} A_P \ A_Q \ A_R ).  
\label{eq:K}
\end{equation}
Hence  $P^M$ satisfies the topological conservation law
$\partial_M P^M \equiv 0$.

As a final note we can also consider the monopole solution in the gauge with vanishing
background gauge field, i.e. $\langle A_5 \rangle \equiv 0$.
We find (for $0 < y < \pi R$)
\begin{equation}
A_5/e \equiv \tilde \Phi = \frac{F(r) - 1}{2 R e} \ (\hat{\roarrow{r}} \cdot \roarrow{T}) \; , \\
\label{eq:ansatz5}
\end{equation}
\begin{equation}
A_{i} = - \frac{G(r) - 1 }{r} \; (\roarrow{T} \times \hat{\roarrow{r}})_i \ \cos\frac{y}{2 R} + \frac{G(r) - 1}{r} \; (T_i - \hat r_i (\hat{\roarrow{r}} \cdot \roarrow{T})) \ \sin\frac{y}{2 R} - \frac{1}{r} \ (\roarrow{T} \times \hat{\roarrow{r}})_i .
\label{eq:ansatz6}
\end{equation} 
Then for $- \pi R < y < 0$ we obtain, by explicitly gauge transforming the fields in Eqs. (\ref{eq:ansatz3}) and (\ref{eq:ansatz4}), $A_5(-y) = - A_5(y)$
and $A_i(-y) = A_i(y)$ as expected from ``naive" parity,
Eqs. (\ref{eq:parity}) and (\ref{eq:parity5}).

\section{Conclusions \label{sec:conclusion}}
In this paper we have discussed Wilson loop symmetry breaking on orbifolds
in five dimensions.   We have tried to make the discussion of symmetry
breaking on orbifolds self contained.  We have also cleared up, in our
minds, the mathematical correspondence between $S^1/Z_2$ orbifolds
with a background gauge field and $S^1/(Z_2 \times Z_2^\prime)$ orbifolds which have been considered in the literature.  In
fact they are identical upon rescaling the radius by a factor of 2.  
Although our analysis has been in non-supersymmetric gauge theories,
it should be easy to extend to the case of orbifold symmetry breaking
in supersymmetric gauge theories.

We have constructed monopole string solutions for an
$SO(3)$ gauge group; valid when $SO(3)$ is broken to $U(1)$.
Our construction can be extended to any $SU(N)$ gauge group on an 
$M \times S^1/Z_2$ orbifold with background gauge field.   Such monopole strings may have interesting phenomenological consequences for grand unified scenarios with large extra dimensions \cite{guts}.  They would be expected to have mass of order $1/ 2 \alpha R$, with a compactification scale $1/R$ as small as a few TeV.  Note that a GUT monopole string can lead to catalysis of baryon number violating processes \cite{catalysis}.

Another interesting example would be in the case of the $SU(3)$ electroweak unification model recently discussed in the literature
\cite{su3_unification}.  It is easy to show that this model also contains
monopole strings when the symmetry is broken to either $SU(2) \times U(1)_Y$ or directly to $U(1)_{EM}$ with the addition of a Higgs multiplet in the triplet
representation.   Such a monopole string will have mass of order $60/R$.   

Clearly in light of the results presented here, it will be interesting to study 
monopole string production at high energy accelerators and at 
finite temperatures in the early universe.   Tree level monopole string 
pair production cross sections at high energy colliders
can be obtained from the usual Drell-Yan formulae,  renormalized by the
coupling of the monopole string,  $e \rightarrow 1/e$, ignoring any form factor  effects \footnote{Tree level cross sections can be suitably ``unitarized" to satisfy unitarity.}.  If the monopoles are light enough, they can be
pair produced in existing colliders such as the Fermilab Tevatron, and
trapped and bound in the matter surrounding the collision region.  
Current bounds, using the samples exposed in the D0 experiment, on the production cross section of such monopole pairs is $0.42$ pb, yielding
a monopole mass limit of about 355 GeV \cite{kalbfleisch}. 
LHC should be able to push the limits to several TeV, or observe signals.

The mass of the monopole string, discussed in this paper, depends on the
orbifold compactification scale.  Current collider bounds on the compactification scale depend on the scenario for the Standard Model 
particles.  If only gauge fields propagate in the extra dimension,
$y$, then the bound on $1/R$ is about a TeV; hence the mass of the monopole string can be about 60 TeV or higher. If all the SM particles propagate 
in the extra dimensions (the case of so-called Universal Extra Dimensions),  
the collider bound is significantly lower, around 350 GeV
\cite{universal}. In this case, the monopole string could be as light as about 20 TeV.  Such monopole pairs will be copiously produced in future very high energy hadron colliders, such as the proposed VLHC, via Drell-Yan pair production.  

 Finally we expect that at temperatures much
above the compactification scale,  the thermal averaged holonomy will vanish
and the monopole solutions will cease to be relevant.  Then as the universe
cools below the compactification scale, the universe is expected to go
through a symmetry breaking phase transition in which monopoles will be
produced via the Kibble mechanism with roughly one monopole per horizon
size at the transition temperature \cite{kibble}.  Of course any constraints from bounds on cosmological monopole fluxes will require detailed finite temperature analyses.  There are very stringent limits on monopole fluxes, both laboratory and galactic.  The galactic limit on the monopole flux ($F$) is  $F < 10^{-15}$/cm$^2$/sec/str, the so called ``Parker Limit" \cite{parker}. 
A monopole with mass less than about $10^7$ TeV will satisfy the
Parker Limit.  Also, such a relativistic monopole will not over close the
universe. It has been suggested \cite{kephart} that such monopoles may be 
the cosmic ray primary responsible for producing the ultra high energy cosmic rays \cite{uhecr} 
observed on earth going beyond the GZK cut off \cite{GZK}.  The mass of  
the monopole string solutions, presented in this paper, is determined by the compactification scale which could be around $10^7$ TeV.  Thus these monopole
strings could be a candidate for the above scenario.  The laboratory bound 
for the flux of monopoles of any mass and with any velocity passing through the Earth's surface is $7.2 \times 10^{-13}$/cm$^2$/sec/str \cite{flux_bounds}.

\begin{acknowledgments}
S.R. and R.D. would like to thank S. Mathur and C. Imbimbo for discussions. 
R.D. and S.R. received partial support for this work from DOE contract DOE/ER/01545-831. The work of SN was supported in part by US DOE Grants DE-FG030-98ER-41076 and DE-FG-02-01ER-45684.
\end{acknowledgments}


\end{document}

%% file: line_circle.pstex_t
\begin{picture}(0,0)%
\includegraphics{line_circle.pstex}%
\end{picture}%
\setlength{\unitlength}{2960sp}%
\begingroup\makeatletter\ifx\SetFigFont\undefined
\def\x#1#2#3#4#5#6#7\relax{\def\x{#1#2#3#4#5#6}}%
\expandafter\x\fmtname xxxxxx\relax \def\y{splain}%
\ifx\x\y   
\gdef\SetFigFont#1#2#3{%
  \ifnum #1<17\tiny\else \ifnum #1<20\small\else
  \ifnum #1<24\normalsize\else \ifnum #1<29\large\else
  \ifnum #1<34\Large\else \ifnum #1<41\LARGE\else
     \huge\fi\fi\fi\fi\fi\fi
  \csname #3\endcsname}%
\else
\gdef\SetFigFont#1#2#3{\begingroup
  \count@#1\relax \ifnum 25<\count@\count@25\fi
  \def\x{\endgroup\@setsize\SetFigFont{#2pt}}%
  \expandafter\x
    \csname \romannumeral\the\count@ pt\expandafter\endcsname
    \csname @\romannumeral\the\count@ pt\endcsname
  \csname #3\endcsname}%
\fi
\fi\endgroup
\begin{picture}(10062,3070)(1789,-2519)
\put(2401,-1336){\makebox(0,0)[b]{\smash{\SetFigFont{9}{10.8}{rm}{\color[rgb]{0,0,0}-2$\pi$ R}%
}}}
\put(3601,-1336){\makebox(0,0)[b]{\smash{\SetFigFont{9}{10.8}{rm}{\color[rgb]{0,0,0}-$\pi$ R}%
}}}
\put(4801,-1336){\makebox(0,0)[b]{\smash{\SetFigFont{9}{10.8}{rm}{\color[rgb]{0,0,0}0}%
}}}
\put(6001,-1336){\makebox(0,0)[b]{\smash{\SetFigFont{9}{10.8}{rm}{\color[rgb]{0,0,0}$\pi$ R}%
}}}
\put(7201,-1336){\makebox(0,0)[b]{\smash{\SetFigFont{9}{10.8}{rm}{\color[rgb]{0,0,0}2$\pi$ R}%
}}}
\put(4801,-61){\makebox(0,0)[b]{\smash{\SetFigFont{12}{14.4}{rm}{\color[rgb]{0,0,1}$Z_2$}%
}}}
\put(3976,-661){\makebox(0,0)[b]{\smash{\SetFigFont{12}{14.4}{rm}{\color[rgb]{0,0,1}$\cal T$}%
}}}
\put(6001,-2011){\makebox(0,0)[b]{\smash{\SetFigFont{12}{14.4}{rm}{\color[rgb]{0,1,0}$Z_2^\prime$}%
}}}
\put(10351,389){\makebox(0,0)[b]{\smash{\SetFigFont{9}{10.8}{rm}{\color[rgb]{0,0,0}0}%
}}}
\put(11851,-1261){\makebox(0,0)[b]{\smash{\SetFigFont{9}{10.8}{rm}{\color[rgb]{0,0,0}$\pi$ R}%
}}}
\put(10651,-2461){\makebox(0,0)[b]{\smash{\SetFigFont{9}{10.8}{rm}{\color[rgb]{0,0,0}2$\pi$ R}%
}}}
\put(8551,-1261){\makebox(0,0)[b]{\smash{\SetFigFont{9}{10.8}{rm}{\color[rgb]{0,0,0}-$\pi$ R}%
}}}
\put(10201,-1711){\makebox(0,0)[b]{\smash{\SetFigFont{12}{14.4}{rm}{\color[rgb]{0,0,1}$Z_2$}%
}}}
\put(9976,-886){\makebox(0,0)[b]{\smash{\SetFigFont{12}{14.4}{rm}{\color[rgb]{0,0,1}$\cal T$}%
}}}
\put(11251,-1036){\makebox(0,0)[b]{\smash{\SetFigFont{12}{14.4}{rm}{\color[rgb]{0,1,0}$Z_2^\prime$}%
}}}
\end{picture}